\documentclass[
amsmath,amssymb,
physrev,
twocolumn 
]{revtex4-2}

\usepackage{graphicx}
\usepackage{dcolumn}
\usepackage{bm}

\usepackage{color}

\newcommand{\leri}[1]{\left(#1\right)}

\begin{document}


\title{\textbf{Reproducing anomalous transport coefficients from electro-static tokamak edge turbulent dynamics} 
}%

\author{Fabio Moretti$^{1}$}
\author{Francesco Cianfrani$^{1}$}
\author{Nakia Carlevaro$^{1}$}
\author{Giovanni Montani$^{1,2}$}

\affiliation{%
$^{1}$Nuclear Department, ENEA, C.R. Frascati, Via E. Fermi 45, 00044 Frascati (Roma), Italy\\
$^{2}$Physics Department, ``Sapienza'' University of Rome, P.le Aldo Moro 5, (00185) Roma, Italy
}%



\begin{abstract}
Turbulent transport near the X-point of a large tokamak is examined using local, gradient-driven simulations that determine the saturated plasma profiles. The distribution of a representative set of particle tracers evolving within these profiles is then analyzed. The study demonstrates that the resulting transport is diffusive, characterized by a coefficient that depends on the spectral properties of the turbulent energy and attains anomalous high values under broad conditions. These findings suggest that anomalous transport is an inherent outcome of the fundamental non-linear drift dynamics of plasmas. The scaling of transport with turbulent energy is also addressed, with implications for future progress toward a mean-field framework for turbulent transport.    
\end{abstract}

\maketitle

\textit{Introduction}\textemdash Tokamak devices are among the most extensively investigated and, arguably, the most promising candidates for the realization of a prototype nuclear fusion reactor \cite{2024Natur.629..555D,PhysRevLett.134.185102}. One of the principal challenges in achieving good plasma confinement in a tokamak arises from the so-called anomalous transport process \cite{Haas1986ConceptualAE,Thoul1994AnomalousPT,KRUPIN_2024}, namely an outward flux of heat and particles that exceeds the standard predictions based on the two main regimes of plasma collisionality, i.e. the Braginskii classical \cite{Braginskii:1965} and neoclassical  \cite{1968JETP...26..233G,RevModPhys.48.239} ones.

The dominant mechanism underlying this anomalous transport is generally attributed to drift turbulence, which produces a strong coupling between density and vorticity
\cite{hase-waka83,hase-waka87}. In this framework, the vorticity field is given by the perpendicular Laplacian of the electric potential with respect to the background magnetic field configuration, and the advection by the electric potential itself constitutes the fundamental non-linear ingredient driving turbulent cascades \cite{scott02}. Pressure fluctuation behaves as a passively advected scalar, with the parallel current divergence inducing a non-trivial coupling with the vorticity equation and background gradients triggering a free-energy source. It is important to emphasize that the so-called non-linear drift response does not arise solely from the saturation of a linear instability —such as that associated with drift waves— but rather represents an intrinsically self-sustained process \cite{scott90,Scott_2007,montani-fluids2022,2023PhyD..45133774M,sym15091745}.

Although the connection between electrostatic drift turbulence and enhanced particle transport coefficients is widely accepted, a precise quantitative mapping between the diffusion coefficient appearing in two-fluid models and the effective particle diffusion coefficient remains insufficiently characterized. Furthermore, the role of the specific magnetic geometry is not yet fully understood. Electrostatic drift turbulence is not the only mechanism capable of triggering turbulent particle transport; other relevant instabilities include, among others, Ion Temperature Gradient modes \cite{1983PhFl...26..673G}, trapped electron modes \cite{PhysRevLett.95.085001} and the so-called interchange instability \cite{PhysRevLett.100.225002}, driven by magnetic field curvature.

Recent advances in turbulence simulation codes for tokamak plasmas (see the review \cite{SCHWANDER2024106141}, and in particular  \cite{HALPERN2016388,2019PhPl...26e2517S,Bufferand_2021}) have demonstrated good agreement with experimental observations at the machine mid-plane. However, these models still face significant difficulties in accurately predicting turbulence properties near the divertor region \cite{2022NucFu..62i6001O}, where the largest fraction of heat power is transported outward.

In this work, we investigate an electrostatic turbulence scenario in the vicinity of an X-point magnetic configuration for a large-size tokamak such as DTT \cite{dtt19}. We characterize in detail the spectral properties of turbulent saturated states and subsequently analyze the evolution of passive tracers advected by the $\mathrm{\textbf{E}\times \textbf{B}}$ plasma flow \cite{2023JPlPh..89a9008S}.
Considering that intrinsic (i.e. pre-turbulent) transport processes in the tokamak edge lie between the Braginskii classical and neoclassical regimes—owing to higher collisionality than in the core, yet still influenced by magnetic geometry— we examine both regimes separately. For each case, we explore significantly different values of the background pressure gradient, corresponding to different strengths of the linear instability drive.

The turbulence dynamics and the associated tracer evolution are studied under periodic boundary conditions and a well-defined diffusive transport clearly emerges (see \cite{PhysRevLett.132.205101} for a comparison with gyro-kinetic simulations). In this regime, we estimate the effective particle diffusion coefficient and analyze its dependence on the turbulent energy content for the different scenarios. Our analysis yields two principal results: \textit{i)} independently of whether Braginskii or neoclassical intrinsic diffusion coefficients are adopted, the resulting effective particle diffusion coefficient consistently reproduces the order of magnitude of anomalous outward transport; \textit{ii)} the scaling of the diffusion coefficient with the turbulent energy is close to a square-root dependence typical of a passively advected scalar in Reynolds-averaged Navier-Stokes (RANS) fluid model with gradient-diffusion hypothesis \cite{Pope_2000}.

These findings lead to two important conclusions. First, since the reduced turbulence model of the tokamak edge here considered successfully captures the essential features of anomalous transport, this strongly suggests that the non-linear advection and the coupling of plasma pressure and electric vorticity constitute the key physical mechanism governing the anomalous regime. Second, the observation that the transport coefficient scales with turbulent energy consistently with RANS neutral fluid turbulence indicates that although non-axial modes play a significant role in energy transfer and shape the spectral turbulent profiles \cite{biskamp95,CIANFRANI2025134800,CIANFRANI2025134851}, they are not decisive in determining the overall transport process. In this respect, plasma transport at the edge does not differ substantially from that of 2D Eulerian neutral fluid turbulence.

\textit{Methodology}\textemdash Low-frequency plasma drift turbulence in the tokamak edge is well described by a two-fluid model (ions and electrons) with the drift ordering approximation \cite{10.1002/ctpp.200410012}. We consider a local poloidal region close to the X-point that fully extends in the  toroidal direction. We use Cartesian coordinates $(x,y,z)$ (diamagnetic and curvature contributions are neglected) and the background magnetic field $\textbf{B}$ is assumed as
\begin{equation}
\textbf{B}=B \;\!\!\hat{\,\,\textbf{b}}=B_p y\hat{\textbf{e}}_x+B_p x\hat{\textbf{e}}_y+B_t\hat{\textbf{e}}_z\, , 
	\label{lm}
\end{equation}
where the constants $B_p$ and $B_t$ are the poloidal and toroidal magnetic contributions, respectively. In this work, we study the electrostatic limit in a gradient driven approach: the (dimensionless) dynamical variables are thus the electric potential $\phi$ and the pressure perturbation $p$. The model equations read
\begin{align}
    & \partial_t\nabla^2_\perp \phi +\frac{c_s^2}{\Omega_i}\left \{ \phi, \nabla^2_\perp \phi \right \}=c_1\nabla^2_\parallel \leri{p-\phi}+\nu \nabla^4_\perp \phi\;, \label{syst1}\\
    &\partial_t p +\frac{c_s^2}{\Omega_i}\left \{ \phi, p \right \}=-\frac{c_s^2}{\Omega_i\ell_0} \partial_y \phi+c_2\nabla^2_\parallel \leri{p-\phi}+\chi_\perp \nabla^2_\perp p,\label{syst2}
\end{align}
where $t$ is time and $\{A,B\}\simeq \partial_yA\partial_x B-\partial_xA\partial_yB$ denotes the Poisson brackets, while $\nabla_{\parallel}=\!\!\hat{\,\,\textbf{b}}(\!\!\hat{\,\,\textbf{b}}\cdot\nabla)$ and $\nabla_{\perp}=\nabla-\nabla_\parallel$ (see \cite{CIANFRANI2025134800} for more details). The term proportional to $\partial_y\phi$ in Eq.\eqref{syst2} represents the source of turbulent energy provided by the background pressure gradient and the parameter $\ell_0$ the associated length scale. The constants are $c_1=v_A^2/\eta $ and $c_2=\lambda_D^2c^2/\eta$ where $c_s$, $v_A$ and $\lambda_D$ are sound speed, Alfvén velocity and Debye length, while $\chi_\perp$, $\nu$ and $\eta$ denote the diffusivity, viscosity and resistivity parameters, respectively. 

Simulations are run assuming typical tokamak parameters of incoming devices, like DTT \cite{dtt19}: $ n_0 = 5\times10^{19}\, \text{m}^{-3}, \, T=100 \, \text{eV},$ $B_t=3\, \text{T}, \, R=2.1 \, \text{m}, \, a=0.7\, \text{m}$. The poloidal domain is a square of side $2\, \text{cm}$ and the simulation time amounts to $0.35\, \text{ms}$. 
The viscosity and resistivity are kept fixed to $\nu=0.0024\, \text{m}^{2}/\text{s}$ and $\eta=0.88\, \text{m}^{2}/\text{s}$ and we investigate two distinct scenarios: in the first (classical), we assume the Braginskii value $\chi_\perp= 0.011\, \text{m}^2/\text{s}$, while in the second a ten-times-larger neoclassical value $\chi_\perp= 0.11\, \text{m}^2/\text{s}$ is considered. For each of these setups, we test five values for the parameter $\ell_0$, i.e. $\ell_0=2,\, 3.5,\, 5, \, 7.5,\,10\,\text{cm}$ (A, B, C, D, E). 
The dynamical system is solved assuming periodic boundary conditions on the spatial domain (see \cite{CIANFRANI2025134800,CIANFRANI2025134851}) and run using a pseudo-spectral scheme with $47\times 47$ poloidal modes and $11$ toroidal ones. This choice safely complies the physical cut-off related to the Larmor radius ($\rho_i=0.5\,\text{mm}$) in the fluid description.
\begin{figure}[hbtp!]
\centering
\includegraphics[width=8cm]{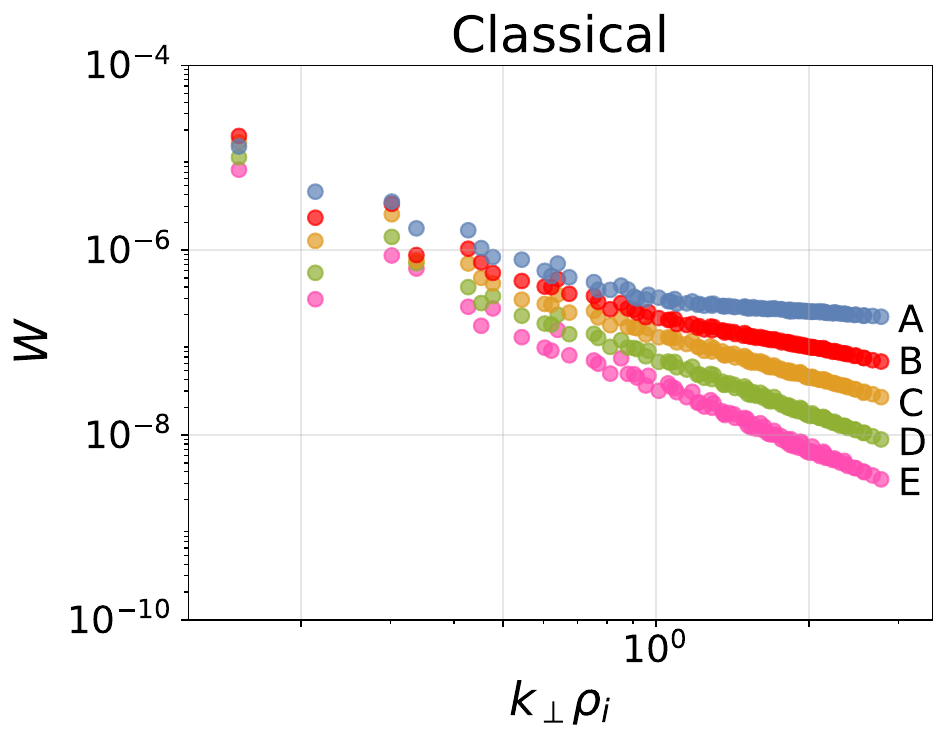}
\includegraphics[width=8cm]{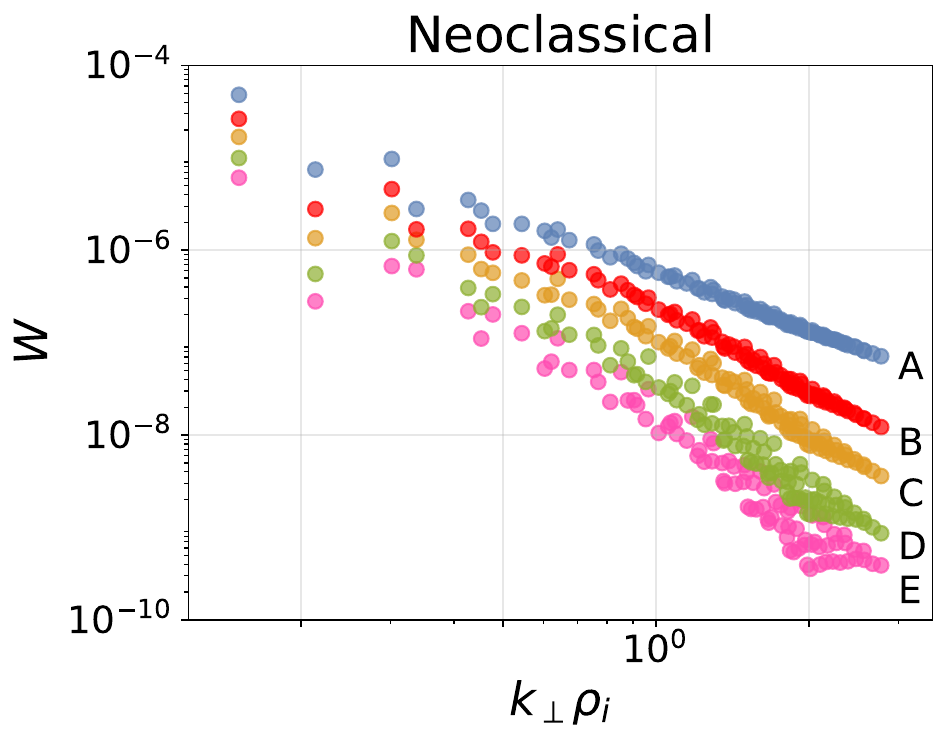}
\caption{Plots of the asymptotic energy spectral density $W$ as a function of the perpendicular wavenumber $k_\perp$.}
\label{fig:spectra}
\end{figure}

The characteristics of the simulated turbulent scenarios are summarized in Fig. \ref{fig:spectra}, where we show the energy spectral density $W=\frac{1}{2}(\nabla_\bot \phi)^2$. It is worth noting how higher more gentle profiles, thus more energetic short wave-length tails, are obtained by increasing the turbulent free energy source due to background pressure gradients (case E $\to$ A) and/or changing diffusivity from neoclassical to classical values.

In what follows, we will discuss how these different spectral features impact on the turbulent transport.
We evolve 5000 test particles, which trajectories are calculated from $\partial_t\mathbf{x}= \mathbf{v}_E=(c/B^2)\,\mathbf{B}\times \mathbf{\nabla}\phi$, $\mathbf{x}$ being the tracer position vector.
The statistical analysis is implemented by calculating the mean squared displacement $\text{MSD}(t)=\langle  \leri{x(t)-x(0)}^2 +\leri{y(t)-y(0)}^2+\leri{z(t)-z(0)}^2 \rangle$, here $\langle \cdot \rangle$ represents an average over the tracer ensemble.

\begin{figure}[hbtp!]
\centering
\includegraphics[width=8cm]{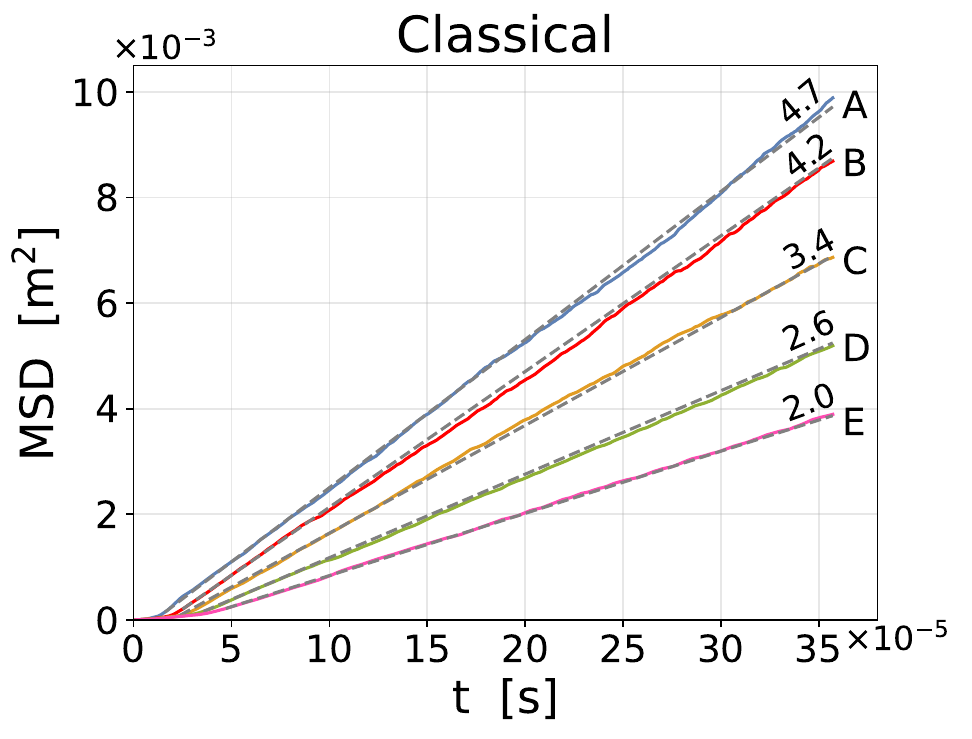}
\includegraphics[width=8cm]{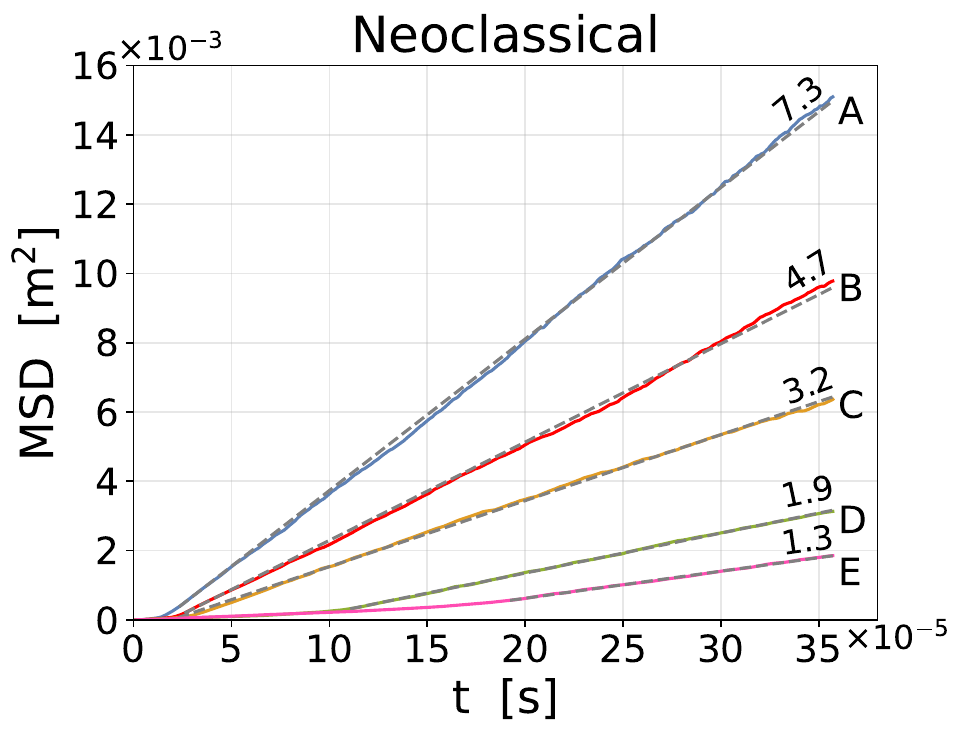}
\caption{MSD plots: a linear fit is superimposed on each curve, with the corresponding diffusion coefficient in $\text{m}^2/\text{s}$  shown on top.}
\label{fig:MSD}
\end{figure}

\textit{Results}\textemdash Let us start by analyzing the evolution of the MSD for the cases under investigation. It is well known that a linear profile is connected to a diffusive phase and, when this occurs, a turbulent  diffusion coefficient $\mathcal{D}_T$ can be defined as $1/6$ (owing to dimensionality) of the MSD slope. The results of this analysis are summarized in Fig.\ref{fig:MSD}. In all the cases considered, we observe that, after an initial transient phase, the system undergoes a stationary diffusive regime. 
\noindent The diffusion coefficients that we estimate from this linear phase are shown as labels. They increase with a stronger free-energy source (E$\,\to$ A) as expected, while the comparison between classical and neoclassical diffusivity is less trivial. In fact, $\mathcal{D}_{T}$ is larger for classical diffusivity in all cases but A and B. This result for C, D and E can be understood owing to high energetic short wave-length tails, while transport in cases A and B is mostly due to the long and intermediate modes ($k_\bot\lesssim 5$) that have higher spectra for neoclassical than for classical diffusivity (see Fig.\ref{fig:spectra}).

Generically, the diffusion coefficients roughly span from $1$ to $8 \, \text{m}^2/ \text{s}$. This implies, for both classical and neoclassical scenarios, an effective transport of the same magnitude as the anomalous transport observed in magnetically confined nuclear fusion experiments \cite{Haas1986ConceptualAE,Thoul1994AnomalousPT,KRUPIN_2024}. Our analysis thus confirms the general assumption that the anomalous transport component is self-consistently generated by the turbulence dynamics of micro-instabilities (at scales $\lesssim 1$ cm). It is worth noting that this result holds regardless of the choice of the bare coefficient $\chi_\perp$.
\begin{figure}[hbtp!]
\centering
\includegraphics[width=8cm]
{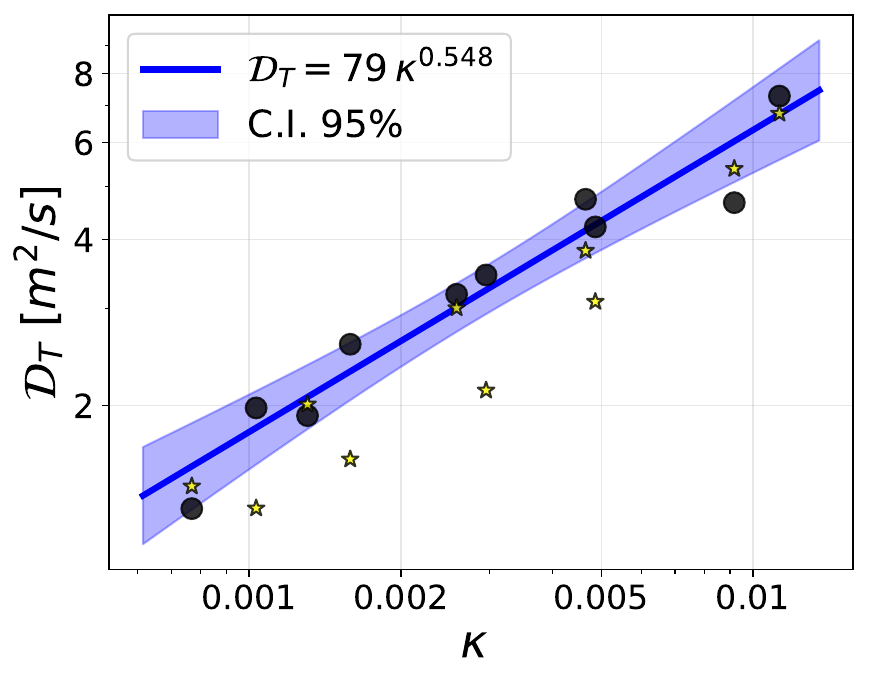}
\caption{Power law scaling of the form $\mathcal{D}_T\propto \kappa^\gamma$. Black dots represent the diffusion coefficients calculated from the tracers, yellow stars the estimates from the QL theory $\mathcal{D}^{QL}_{T}$.}
\label{fig:scalinglaw}
\end{figure}

In view of constructing a mean-field reduced model of edge tokamak turbulence, similarly to the RANS approach \cite{Pope_2000}, we now discuss how the effective diffusion coefficient depends on the associated turbulent energy density. The latter is calculated as the temporal and spatial average at saturation of the electric field energy density normalized with the sound speed $c_s\simeq 7 \times 10^4$ m/s, i.e. $\kappa= \frac{c^2}{2B_t^2c_s^2}\left <|\nabla_\perp \phi|^2 \right >$. We fit the diffusion coefficients assuming a model $\mathcal{D}_T= C \kappa^\gamma$. The obtained parameters are: $C=79\pm24 \,\text{m}^2/\text{s}$ and $\gamma=0.548\pm0.050$ as shown in Fig. \ref{fig:scalinglaw} where we also outline the 95\% confidence interval region. We first note that the best-fit value for the exponent $\gamma$ is consistent with the scaling law $\mathcal{D}_T\propto \sqrt{\kappa}$, a form usually adopted in RANS for passively advected scalars under the gradient-mixing hypothesis \cite{Pope_2000} (see also \cite{10.1063/5.0024479} in which the same scaling is obtained at tokamak edge for 2D interchange-like turbulence). This scaling law can be useful for estimating the turbulent transport coefficient directly from $\kappa$ without giving all the structure of the underlying turbulent fields (see for instance \cite{BASCHETTI2019200} with the application of the $\kappa$-$\epsilon$ model in tokamaks).

In Fig.\ref{fig:scalinglaw} we also show the quasi-linear (QL) transport coefficients (represented by stars): $\mathcal{D}^{QL}_{T}=\tau_{\mathrm{ac}}\,c_s^2 \kappa$. Here $\tau_{\mathrm{ac}}$ denotes the turbulent auto-correlation time that has been computed as the time at which the auto-correlation function $\langle v_E(t,\textbf{x}) v_E(t+\Delta t,\textbf{x}) \rangle$ drops by a factor $e$, $v_E$ being the velocity from the simulated profiles. It shows that the time-scale of vortex formation is of few $\mu s$ for all the cases considered, with the corresponding Kubo numbers in the range of 4 to 6 $\times 10^{-2}$. Hence, we found that in some cases, corresponding to neoclassical diffusivity, the QL approximation provides a quite good estimate of the actual transport coefficients, while they are underestimated by at most 50\% for classical diffusivity. Generically, the quasi-linear theory captures the correct order of magnitude of the transport coefficient.

\textit{Concluding remarks}\textemdash We presented the results of the simulations for a very basic fluid model of drift plasma turbulence at the tokamak edge on which the tracer dynamics is analyzed. The main simplifying assumptions are: $i)$ locality in the poloidal plane, with a 2 cm square being considered; $ii)$ X-point magnetic field; $iii)$ gradient-driven scenario. Nonetheless, the model is capable of capturing the order of magnitude of anomalous diffusion in tokamaks, i.e. few m$^2$/s. This result suggests that anomalous transport is an inherent characteristic of edge tokamak plasma, arising from the fundamental features of the non-linear drift response.

Anomalous diffusivity is usually taken as an input in turbulent simulations covering the whole tokamak scrape-off-layer in order to model sub-grid diffusion (see for instance \cite{TAMAIN2016606}). Here, thanks to the locality assumption we can model micro-turbulence at scales from few Larmor radii (0.5 mm) to 2 cm and derive the transport coefficient self-consistently, showing also that an actual diffusive phase is obtained. The coefficient is almost insensitive to the  classical or neoclassical bare diffusivity values. Moreover, it scales as the turbulent energy to the $0.55$ power quite close (within statistical uncertainty) to the square root typical of passively advected scalars in RANS. 

The small Kubo numbers ($\mathcal{O}\left(10^{-2}\right)$) indicate that particle scattering by turbulence is predominantly stochastic, implying that QL theory can provide an adequate description. This expectation is supported by the appearance of a diffusive phase in all the cases examined. Nevertheless, agreement between tracer-based and QL transport coefficients is observed only in regimes characterized by neoclassical diffusivity. In contrast, for cases exhibiting classical diffusivity, deviations from the QL predictions reach $\sim 50\%$, most likely as a consequence of the drift response. Lowering the diffusivity to classical levels enhances small-scale and non-axisymmetric structures (see \cite{CIANFRANI2025134800}), which amplifies the drift contributions in Eqs.\eqref{syst1} and \eqref{syst2}. These effects introduce a non-trivial coupling between pressure and vorticity that lies beyond the scope of QL theory, where pressure is treated merely as a passive scalar advected by turbulent flows.

The scaling obtained for the transport coefficient will serve as the foundation for developing a mean-field model of the turbulent flow. Furthermore, future theoretical and computational work will aim to relax—or, where possible, eliminate—some of the assumptions introduced above to enable simulations that more accurately reflect real experimental conditions.


%

\end{document}